\documentclass{aa}  

\usepackage{graphicx,textcomp}

\usepackage{txfonts}

\begin{document} 

   \title{Sub-surface alteration and related change in reflectance spectra of space-weathered materials}

   \author{Kateřina Chrbolková\inst{1,2,3}\and
          Patricie Halodová\inst{4}\and
          Tomáš Kohout\inst{1,3}\and
          Josef Ďurech\inst{2}\and
          Kenichiro Mizohata\inst{5}\and
          Petr Malý\inst{6}\and
          Václav Dědič\inst{7}\and
          Antti Penttilä\inst{8}\and
          František Trojánek\inst{6} \and
          Rajesh Jarugula\inst{4}
          }

   \institute{
                        Department of Geosciences and Geography, PO Box 64, 00014 University of Helsinki, Finland \\ \email{katerina.chrbolkova@helsinki.fi}
                \and
                Astronomical Institute, Faculty of Mathematics and Physics, Charles University, V Holešovi\v ckách 2, 18000, Prague, Czech Republic    
                \and
                Czech Academy of Sciences, Institute of Geology, Rozvojová 269, 16500 Prague, Czech Republic
                \and
                Research Centre Řež, Hlavní 130, 250 68 Husinec -- Řež, Czech Republic
                \and
                Department of Physics, PO Box 43, 00014 University of Helsinki, Finland
                \and
                Department of Chemical Physics and Optics, Faculty of Mathematics and Physics, Charles University, Ke Karlovu 3, 12116 Prague, Czech Republic
                \and
                Institute of Physics, Faculty of Mathematics and Physics, Charles University, Ke Karlovu 5, 12116 Prague, Czech Republic
                \and
            Department of Physics, PO Box 64, 00014 University of Helsinki, Finland 
             }

   \date{Received February 8, 2022; accepted xx xx, 2022}

  \abstract
   {Airless planetary bodies are studied mainly by remote sensing methods. Reflectance spectroscopy is often used to derive their compositions. One of the main complications for the interpretation of reflectance spectra is surface alteration by space weathering caused by irradiation by solar wind and micrometeoroid particles.}
   {We aim to evaluate the damage to the samples from H$^+$ and laser irradiation and relate it to the observed alteration in the spectra.}
   {We used olivine (OL) and pyroxene (OPX) pellets irradiated by 5 keV H$^+$ ions and individual femtosecond laser pulses and measured their visible (VIS) and near-infrared (NIR) spectra. We observed the pellets with scanning and transmission electron microscopy. We studied structural, mineralogical, and chemical modifications in the samples. Finally, we connected the material observations to changes in the reflectance spectra.}
   {In both minerals, H$^+$ irradiation induces partially amorphous sub-surface layers containing small vesicles. In OL pellets, these vesicles are more tightly packed than in OPX ones. Any related spectral change is mainly in the VIS spectral slope.
   Changes due to laser irradiation are mostly dependent on the material's melting temperature.
   Of all the samples, only the laser-irradiated OL contains nanophase Fe particles, which induce detectable spectral slope change throughout the measured spectral range.
   Our results suggest that spectral changes at VIS--NIR wavelengths are mainly dependent on the thickness of (partially) amorphous sub-surface layers. Furthermore, amorphisation smooths micro-roughness, increasing the contribution of volume scattering and absorption over surface scattering.}
   {Soon after exposure to the space environment, the appearance of partially amorphous sub-surface layers results in rapid changes in the VIS spectral slope. In later stages (onset of micrometeoroid bombardment), we expect an emergence of nanoparticles to also mildly affect the NIR spectral slope. An increase in the dimensions of amorphous layers and vesicles in the more space-weathered material will only cause band-depth variation and darkening.}


   \maketitle

\section{Introduction}
Reflectance spectroscopy of planetary surfaces enables the determination of the mineralogy and physical state of airless planetary surfaces \cite[see, for example, ][]{burns_89}. Planetary spectra contain complex information as they are influenced by multiple additional factors, including temperature, the roughness of the surface, or the space weathering state. In this article, we focus on how space weathering influences planetary materials at the microscale and what consequences it has for the spectra. Such knowledge is crucial for the correct interpretation of spectroscopic observations.

Space weathering is caused by two major processes -- solar wind irradiation and micrometeoroid bombardment -- which alter the topmost layers of planetary surfaces \citep{hapke_65, hapke_01, wehner_63}. As a result, the spectra of silicate-rich bodies darken, the spectral contrast decreases, and the visible (VIS) and near-infrared (NIR) slopes increase (reddening of the spectra), as documented, for example, by \cite{hapke_01}, \cite{pieters_00}, and \cite{pieters_16}. Other processes that influence the space weathering state are, for example, galactic radiation, sublimation, and mixing with the material of impactors, but these were not identified as the main drivers of space weathering change in previous studies of the Moon and of near-Earth asteroids \citep{pieters_16}.

Earlier studies reported on the sub-surface changes related to space weathering, especially changes due to solar-wind sputtering and impact melting with associated vaporisation and vapour redeposition. Already in the lunar samples from the Apollo missions, \cite{keller_93, keller_97} observed thin, 200\,nm, amorphous rims with randomly dispersed inclusions of Fe metal, which they identified as the solar wind irradiation products. Since then, so-called nanophase Fe (npFe$^0$) particles have been identified in many laboratory simulations (see, for example, \citealp{fazio_18}, \citealp{weber_20}, and \citealp{wu_17}) and also in samples from asteroid (25143)~Itokawa \citep{noguchi_14b, noguchi_14a} and the Moon \citep{kling_21}. Other abundant types of structures in the altered sub-surface layers are vesicles and blisters, as documented, for example, by \cite{dobrica_16} and \cite{matsumoto_15} for Itokawa particles. 

Because of the sparsity of the natural samples and the slow natural evolution of space weathering, a significant part of our understanding comes from laboratory experiments. The effect of the solar wind is usually simulated by irradiation of the samples by various ions (see, for example, the work of \citealt{brunetto_14}, \citealt{demyk_01}, \citealt{hapke_70}, or \citealt{loeffler_09} for more details). The effect of micrometeoroid impacts is mainly studied using individual or repeated laser pulses of various durations and energies (see, for example, \citealt{sasaki_01a}, \citealt{sasaki_02}, and \citealt{yamada_99} for usage of nanosecond laser pulses or \citealt{fazio_18} and \citealt{chrbolkova_21} for femtosecond laser pulses). Only rarely do experiments include actual impacts of dust particles, such as in \cite{fiege_19}.

This article is a follow-up of \cite{chrbolkova_21}, where we evaluated the difference in the spectra influenced by solar-wind irradiation and micrometeoroid bombardment. This was done through laboratory irradiations of olivine and pyroxene pellets using low-energy 5 keV H$^+$ ions and individual femtosecond laser pulses as in \cite{fazio_18}. We found that there was a difference in the influence of the two types of irradiation on NIR wavelengths; the laser influenced VIS and the NIR range, while ion irradiation had only a mild effect in the NIR wavelength range. However, in general, the spectral evolution was determined more by the original mineralogy of the samples than by the space weathering agent. This result agreed with earlier observations of \cite{sasaki_02}.

In \cite{chrbolkova_21}, we only evaluated the spectral change, but we did not have information on the material changes within the pellets. In this article, we aim to find the connection between the spectral changes and the sub-surface alteration caused by the different space weathering agents. 
\section{Methods}
\subsection{Samples, irradiations, and spectroscopy}
\label{sect-spectr}
We used samples identical to those in \cite{chrbolkova_21}, that is pellets made of a pressed powder (with particle sizes $<106$\,\textmu m) of olivine (forsterite number Fo90), further denoted as OL, and pyroxene (enstatite number En67), further denoted as OPX, on top of a KBr base. We refer the reader to \cite{chrbolkova_21} for more details.

For the microscopical observations, we used four different irradiated pellets from \cite{chrbolkova_21}. Two of them were produced by H$^+$ irradiation, and the other two by femtosecond laser irradiation. In each case, one OL and one OPX pellet were selected. 

The H$^+$-irradiation was done using 2$\times10^{17}$\,H$^+$/cm$^2$ 5\,keV ions. The ion beam covered the whole 13-mm pellet and was oriented perpendicularly to the surface of the pellet. Samples were placed in a vacuum chamber with a pressure of $\approx10^{-7}$\,mbar during the irradiations.

A femtosecond laser was set to shoot individual, spatially separated 800-nm pulses perpendicularly to the surface of the pellet. The duration of one pulse was 100\,fs, and for the OL sample the energy of one pulse corresponded to 1.5\,mJ, while for OPX it was 1.8\,mJ. The laser spot on the surface of the pellet was $\approx50$\,\textmu m in diameter. The pressure in the vacuum chamber was $\approx10^{-4}$\,mbar. For microscopic observations, we selected pellets in which the spatial separation of the individual laser-induced microcraters was the largest possible so that we were sure that the studied crater is not influenced by the neighbouring craters, that is OL irradiated by 1.7\,mJ/cm$^2$ (equivalent crater distance 300\,\textmu m) and OPX irradiated by 1.5\,mJ/cm$^2$ (200\,\textmu m).

To compare the spectral evolution with the observed micro-structures, we used spectra measured in \cite{chrbolkova_21}. The H$^+$-irradiated spectra were measured using an OL-750 automated spectroradiometric measurement system by Gooch \& Housego with polytetrafluoroethylene (VIS) and gold (NIR) standards and an incident angle of 10$^{\circ}$ \citep{chrbolkova_21, penttila_18}. The laser-irradiated spectra were measured using a Vertex 80v from Bruker with an A513/Q variable angle reflection accessory using the Spectralon standard, an incident angle of 0$^{\circ}$, and a collection angle of 30$^{\circ}$. As we only compared relative trends within one dataset, the differences in the experimental set-up do not inhibit further interpretation. 

A straightforward visualisation of the spectral change is the ratio plot of the weathered to fresh material. For H$^+$ irradiation, we used the ratio plot for the 2$\times10^{17}$\,H$^+$/cm$^2$ case, which exactly matched the samples studied here with the microscope. For laser irradiation, we studied individual craters. To obtain a representation of the spectrum of one crater, which is too small to be captured by our spectrometer, we selected a spectrum captured over a surface homogeneously covered by craters that lie tightly next to each other but do not overlap. Based on microscopical observations, the craters are approximately 100\,\textmu m in diameter; we thus selected spectra of the pellet in which the centres of the craters were 100\,\textmu m apart. These spectra were influenced by the unaltered material located in the corners of the squares circumscribed to the craters, but its contribution was smaller than that from the irradiated material.

\subsection{New analyses}
Using the pellets described in the previous section, we carried out the following analyses to study the surface structures related to space weathering: scanning electron microscopy (SEM) using secondary electrons (SEM-SE) for morphology, backscattered electrons (SEM-BSE), and energy-dispersive X-ray analysis (SEM-EDS) for compositional information. For sub-surface structures, we used transmission electron microscopy (TEM) in conventional (CTEM), high-resolution (HR-TEM), and scanning (STEM) modes combined with EDS for compositional information.

\subsubsection{Scanning electron microscopy}
A Tescan Lyra 3GMU Dual-Beam Scanning Electron Microscope equipped with a field emission gun (FEG) and focused Ga ion beam was used for SEM and in situ lift-out of the lamellae dedicated to TEM analysis. The pellet samples were coated with carbon prior to the electron microanalysis to prevent charging.

The samples were at first characterised with top-view SEM and EDS imaging. The SEM operational conditions were as follows: i) surface imaging conditions -- accelerating voltage of 10\,kV, 5--9\,mm working distance, SE or in-beam SE imaging modes; ii) analytical conditions – accelerating voltage of 15\,kV, 9\,mm working distance, SE, and BSE imaging modes. Chemical analysis was carried out using an EDS system composed of an 80\,mm$^2$ X-Max$^{\textrm{N}}$ Oxford Instruments Silicon Drift Detector and Oxford Instruments AZtec acquisition and processing software.

After the SEM characterisation, thin lamellae for subsequent TEM were prepared by focused ion beam milling. To avoid charging and drifting during the ion preparation, the non-conductive pellets were coated with a 100 nm layer of gold. Thin, cross-sectional specimens were extracted from the regions of interest of the pellet samples with the use of an OmniProbe OP400 nanomanipulator and attached to a copper grid for further ion thinning and polishing to electron transparency (<\,100\,nm thickness). The protective Pt strip was deposited on the surface at the area of interest before the milling process to prevent Ga$^{2+}$ ion beam damage of the laser- and H$^+$-irradiated sample surface. The ion beam operational conditions were as follows: accelerating voltage of 30\,kV and beam current of 2\,nA for the trench milling, 500\,pA for cross-section polishing, 160\,pA for further thinning, and 10\,kV and 40\,pA for the final polishing steps to minimise the Ga ion damage of the thin lamellae.

\subsubsection{Transmission electron microscopy}
The CTEM, HR-TEM, and STEM investigations were carried out using a Jeol JEM 2200FS FEG microscope equipped with an Oxford Instruments Ultim Max TEM EDS system, which was used for the acquisition of elemental maps, line scans, point analysis and chemical quantification. The microscope was operated at 200\,kV. TEM images were acquired using conventional imaging mode for the bright field (BF) and HR. In STEM, high-annular dark-field, BF, and dark-field detectors were used. Selected area diffraction patterns and HR images were captured on a TVIPS TemCam-XF416 4k$\times$4k CMOS camera.

Analyses with STEM-EDS were obtained using a spot size of 1\,nm and integration time of 120\,s. Acquisition of the EDS elemental maps had a typical duration of $\approx$\,30\,min. Mapping and line scans were acquired using the AutoLock function for drift correction. The data were further processed using the AZtecTEM software, and deconvolution of the elements was solved using the Tru-Q technology. Factory standards (Oxford Instruments) were used for quantification. Quantitative analysis used background subtraction by filtered least-squares fitting \citep{statham_77}, and Cliff-Lorimer correction procedures were used for quantification. Unwanted peaks of non-sample elements (C, Cu, Co, Ga) were excluded for the purpose of quantitative analysis.

\subsection{SRIM irradiation simulations}
\label{sect-srim}
To support our observations from the H$^+$ irradiations, we conducted The stopping and range of ions matter (SRIM) \citep{ziegler_85} simulations using 1000 5\,keV H$^+$ ions propagating through the OL and OPX materials with compositions matching our samples. We focused on ion penetration depths and also on the distribution of the collision events in the samples to correlate these with the distribution of the disturbances in the samples. 

\section{Results}
\subsection{Surface changes}
\begin{figure}
        \centering
        \includegraphics[width=\columnwidth]{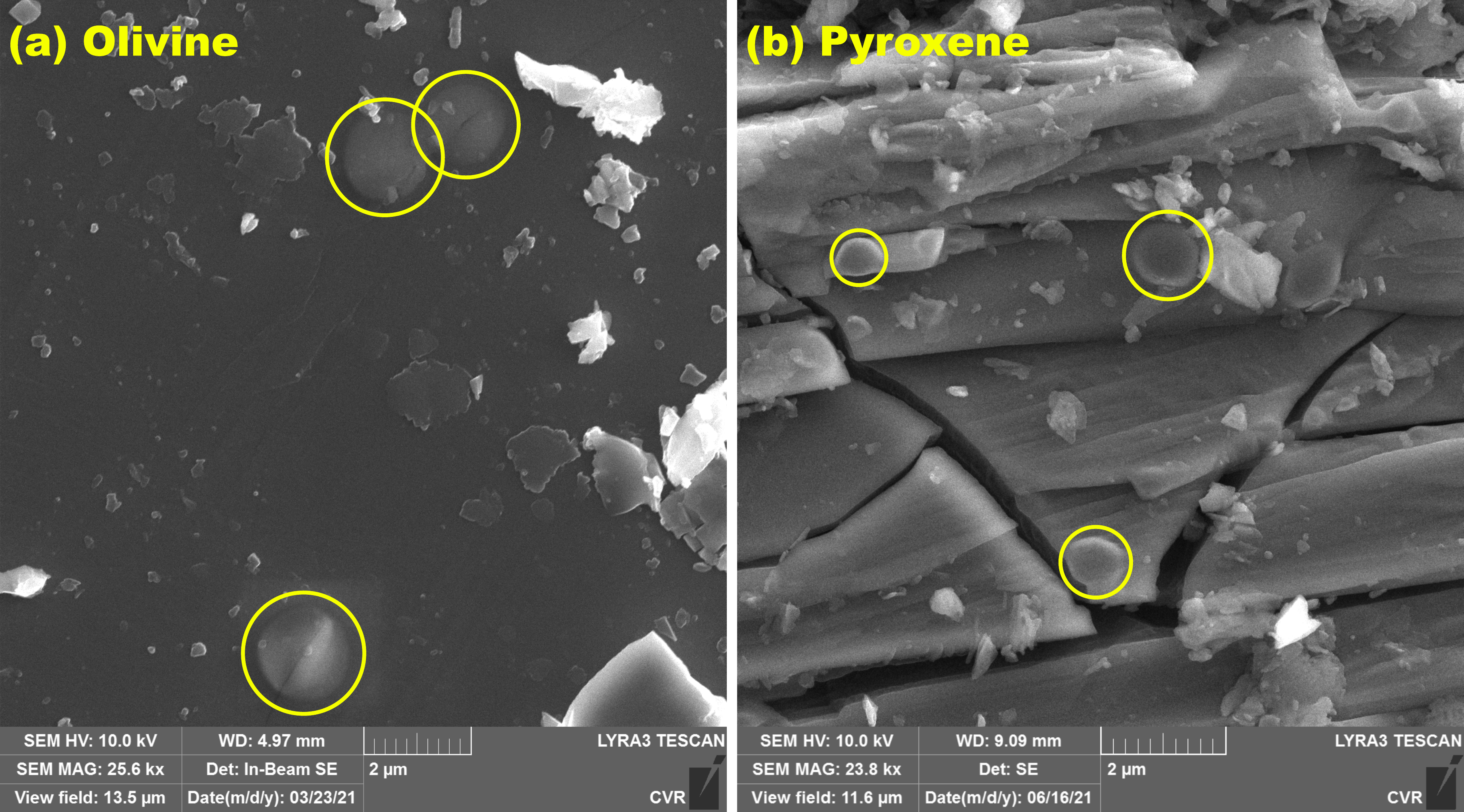}
        \caption{Blisters (highlighted with yellow circles) created by H$^+$ irradiation. Top view of (a) olivine and (b) pyroxene surface captured by scanning electron microscopy. The scale bar in both figures corresponds to 2\,\textmu m.}
    \label{fig-blisters}
\end{figure}

All four samples show visible darkening of their irradiated surfaces. The OL and OPX irradiated by H$^+$ ions do not show major changes in surface morphology. The only visible alteration in the SEM images is in the form of blisters, which are more frequently found in the OPX than in the OL, where the blisters are observed only rarely (see Fig.~\ref{fig-blisters}). We estimated that the size of the blisters is $\approx 1.7$\,\textmu m in the OL, while in the OPX it is $\approx 0.9$\,\textmu m. The morphology of the two surfaces differs, which could have influenced the blister size.

\begin{figure}
        \centering
        \includegraphics[width=\columnwidth]{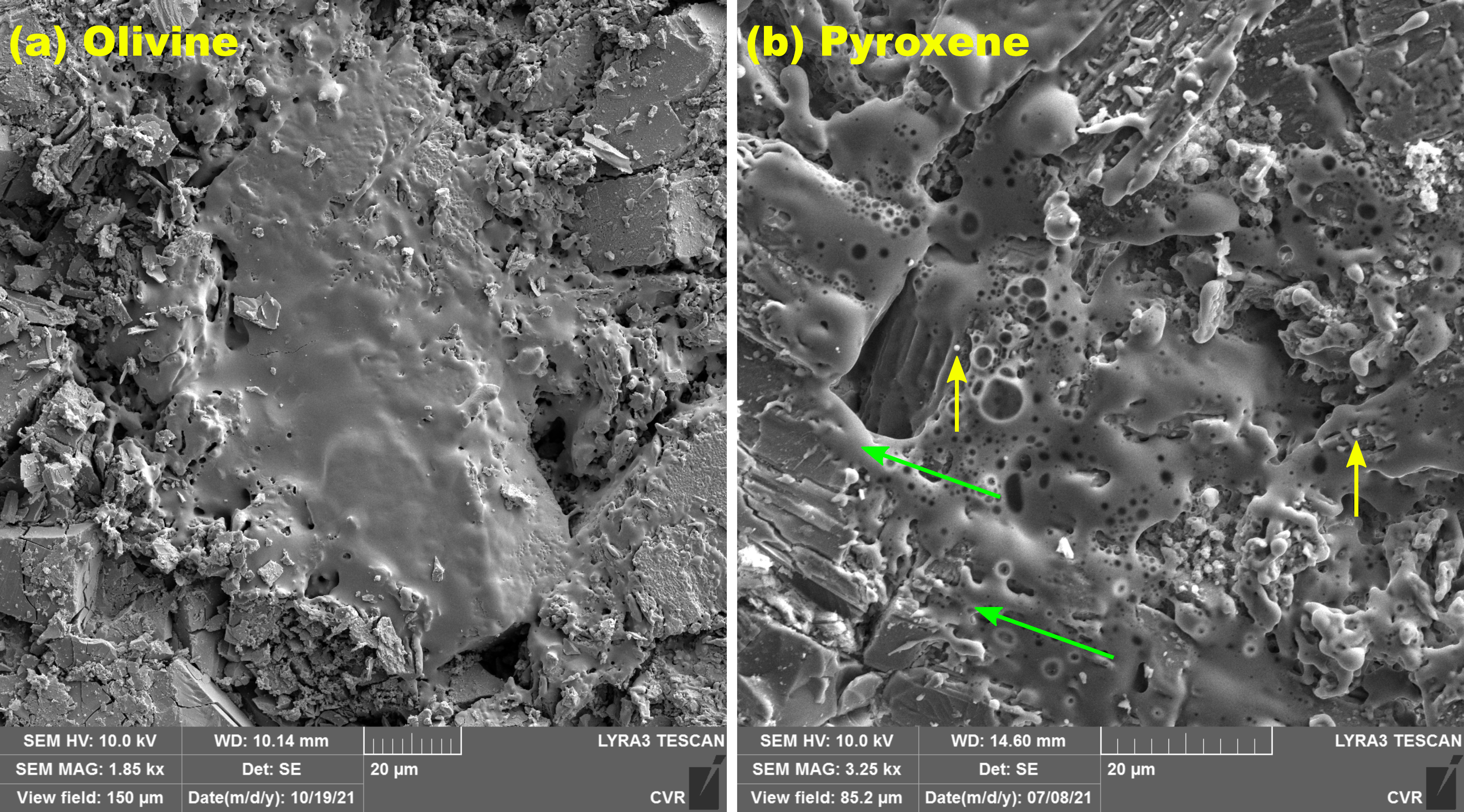}
        \caption{Scanning electron microscopy images (top view) of one individual crater in the (a) olivine and (b) pyroxene sample. The centre of each crater is approximately in the middle of the image. The craters lie in the molten areas in both figures, but marking the border of them would be highly inaccurate. The scale bar in both figures corresponds to 20\,\textmu m. The yellow arrows point to melt drops, and the green arrows show examples of melt splashes.}
        \label{fig-craters}
\end{figure}

Considering the laser-irradiated samples, the surface changes are more extensive, as expected. We observe extensive melting in the area of the crater. In the OL, melting is restricted to the area of the crater and the melt is smooth with a minor amount of bubbles and depressions after gas release. In the OPX, the melting is more pronounced, and we also see melt splashes and melt drops (see Fig.~\ref{fig-craters}). The bubble-rich melted layer often encompasses large continuous areas of the sample surface, and it is more difficult to identify individual craters.  

\begin{figure}
        \centering
        \includegraphics[width=\columnwidth]{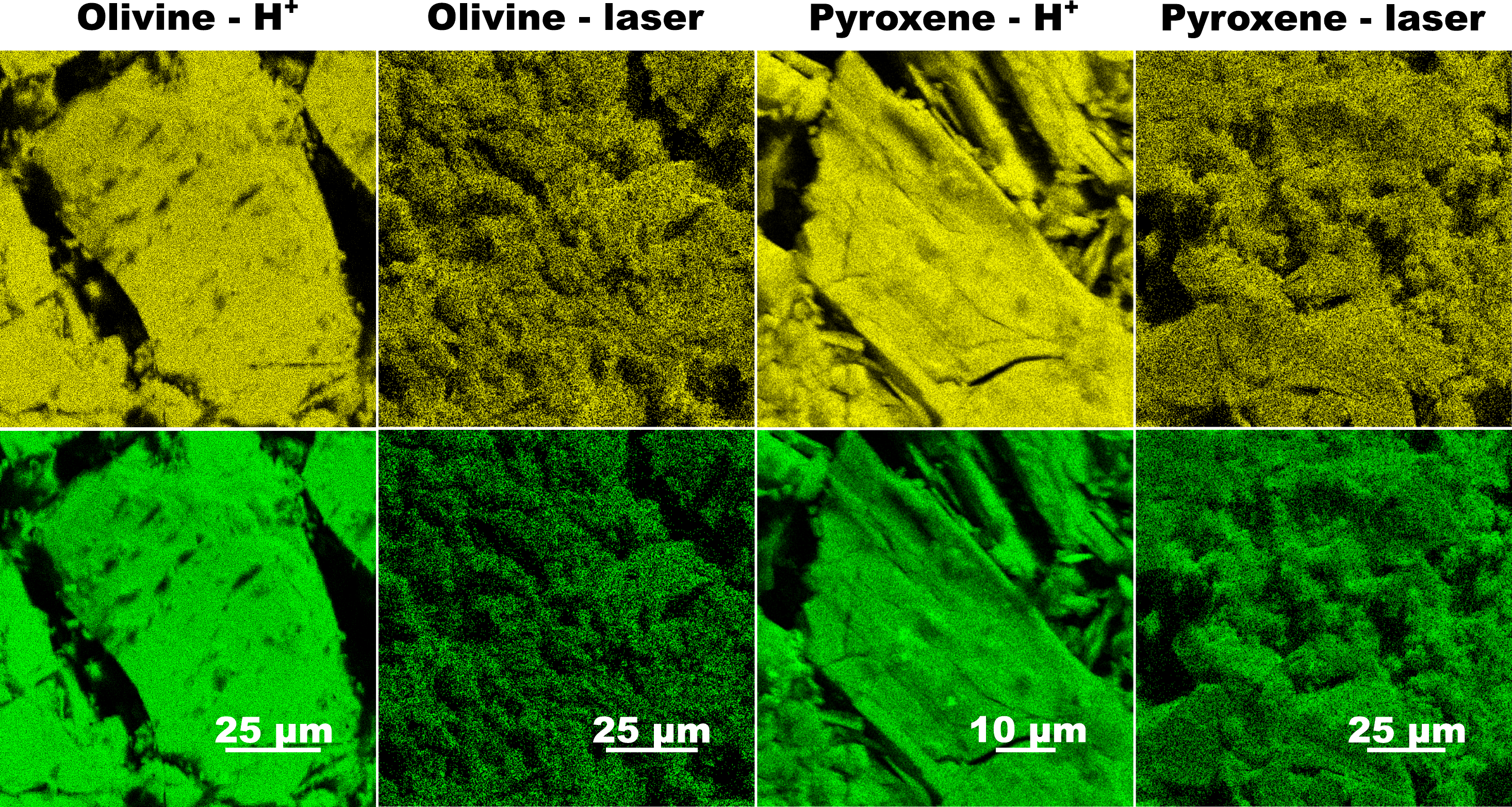}
        \caption{Maps of two selected elements from scanning electron microscopy using energy dispersive X-ray analysis. Si is in yellow (top row), and Mg is in green (bottom row).}
        \label{fig-semchem}
\end{figure}

Maps of irradiated surfaces from SEM-EDS (Fig.~\ref{fig-semchem}) do not show any measurable spatial variation in the composition related to the space weathering experiments. The only visible features are linear structures in one OPX sample related to terrestrial weathering (amphiboles).

\subsection{Sub-surface changes}
\begin{figure}
        \centering
        \includegraphics[width=\columnwidth]{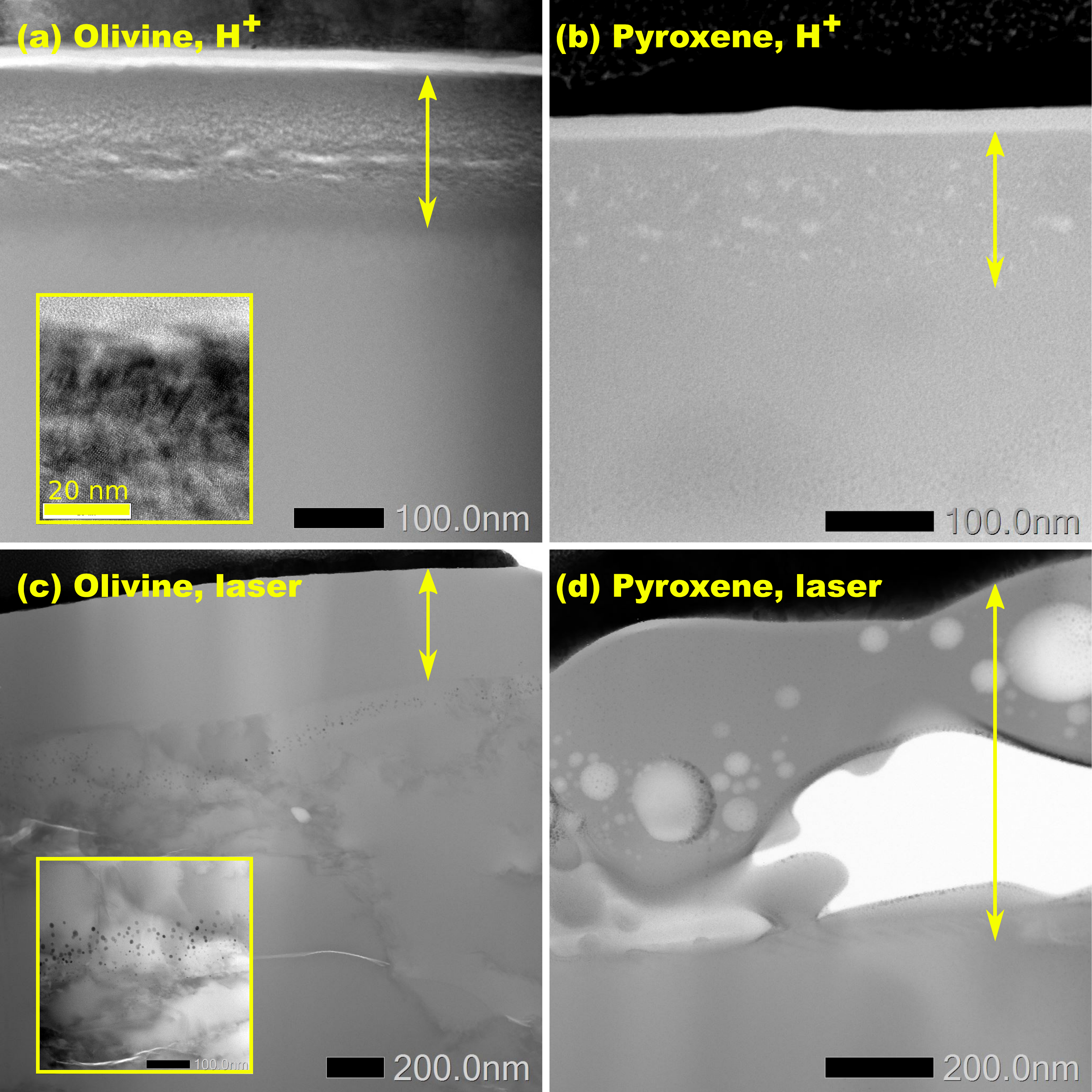}
        \caption{Bright-field scanning transmission electron microscopy images of the lamellae of all samples. Inset in (a) is the high-resolution transmission electron microscopy image of the altered layer in the H$^+$-irradiated sample. Inset in (c) is the zoomed-in picture of the area containing nanophase Fe particles. Double-headed arrows mark the areas that are partially or fully amorphised.}
        \label{fig-TEM}
\end{figure}

The images of H$^+$-irradiated samples from TEM reveal radiation damage with abundant vesicles in thin altered layers. In the OL, the influenced layer is approximately 200\,nm thick, while in the OPX it is approximately 150\,nm thick (see Fig.~\ref{fig-TEM}~(a, b)). The two samples differ also in the size and abundance of the vesicles. In the OL, the vesicles are in multiple, tightly packed layers. Their shape is elongated, with a size of approximately 10$\times$50\,nm (longer side parallel to the surface). In the OPX, the vesicles are more spherical and less abundant, with sizes of approximately 20\,nm.

Diffraction patterns of the sub-surface layers show that the crystalline structure is extremely damaged and in both cases contains fully amorphous areas. We refer to such a structure as a partially amorphous layer. 

Laser-irradiated OL and OPX samples differ significantly from each other. While the OL sample is the only one in which we identify Fe nanoparticles (npFe$^0$, 7--10\,nm in diameter), the OPX sample is characterised by a huge number of vesicles with sizes ranging from several nm to 2\,\textmu m and no nanoparticles (see Fig.~\ref{fig-TEM}~(c, d)). In the OL, the completely amorphous surface layer is approximately 300\,nm thick, and the nanoparticles are directly below it concentrated in a thin irregular zone. The thickness of the fully amorphous layer in the OPX is typically around 400\,nm but it significantly varies and can reach up to 2.5\,\textmu m in some extreme cases. Tiny black dots in the TEM image of the OPX irradiated by laser represent redeposited particles of Pt and Au that we used for imaging and extraction of the thin section for TEM analyses (see Fig.~\ref{fig-TEM}~(d)).

\begin{figure}
        \centering
        \includegraphics[width=\columnwidth]{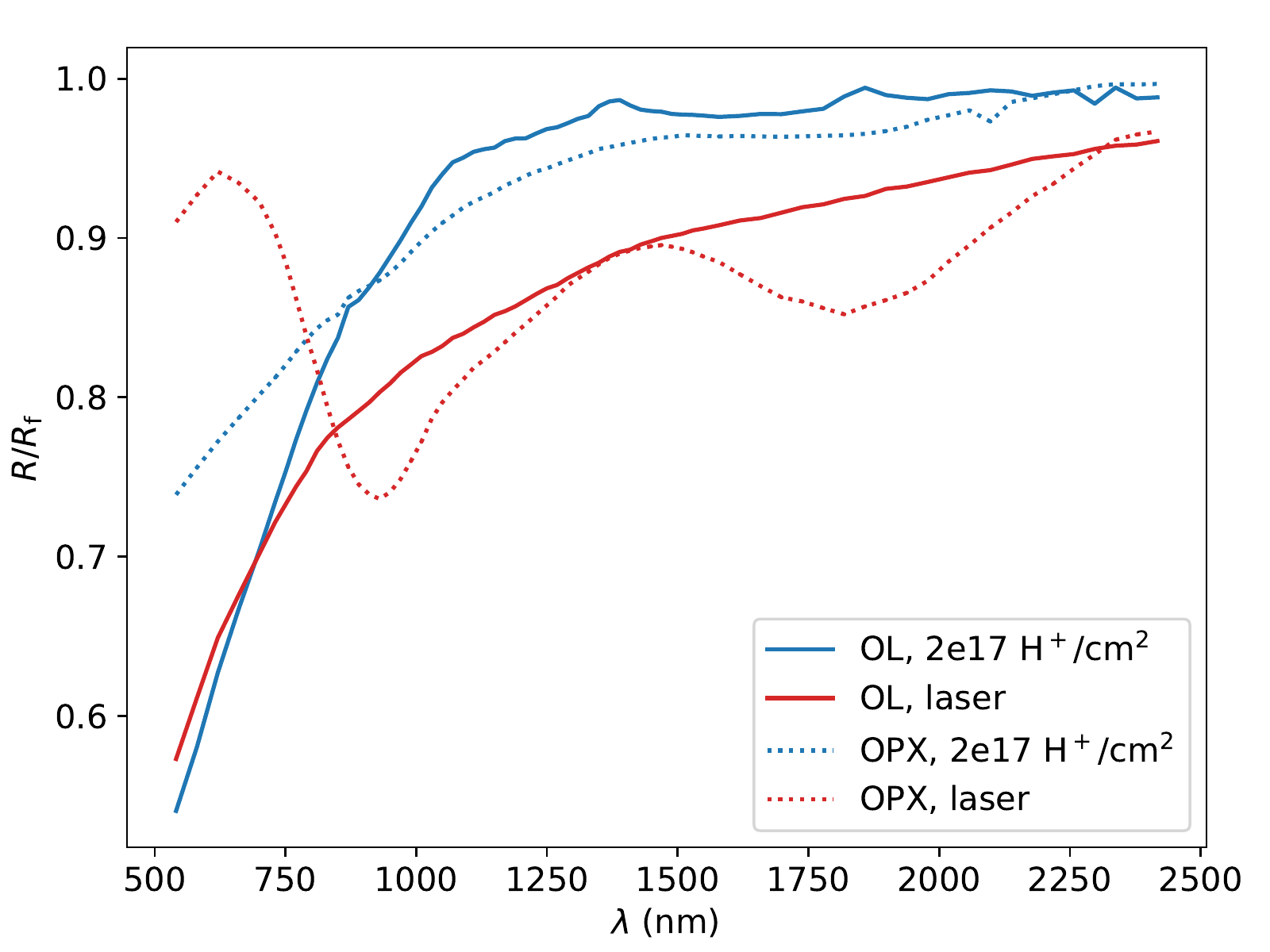}
        \caption{Ratio of mature to fresh spectra of our samples. $R$ stands for the reflectance, $R_f$ for reflectance of the fresh material, and $\lambda$ for wavelength. Laser-irradiated spectra were measured on samples in which the craters were 100\,\textmu m apart. The OPX absorption bands lie at $\approx$930\,nm and 1820\,nm in the un-irradiated spectra; see \citet{chrbolkova_21} for comparison.}
        \label{fig-spectra}
\end{figure}

In Fig.~\ref{fig-spectra}, we show the ratio of irradiated to fresh spectra of the samples, as described in Sect.~\ref{sect-spectr}. Spectral ratios of the OL and OPX irradiated by H$^+$ show similar behaviour, that is a strong decrease of the albedo at shorter wavelengths (resulting in an increase of the spectral slope) and nearly no change at longer wavelengths. The spectrum of the laser-irradiated OL also behaves similarly but shows additional changes in the NIR slope. In contrast, spectra of the laser-irradiated OPX do not show changes in spectral slope, although darkening and enhancement of mineral absorption bands are observed.

\begin{figure}
        \centering
        \includegraphics[width=\columnwidth]{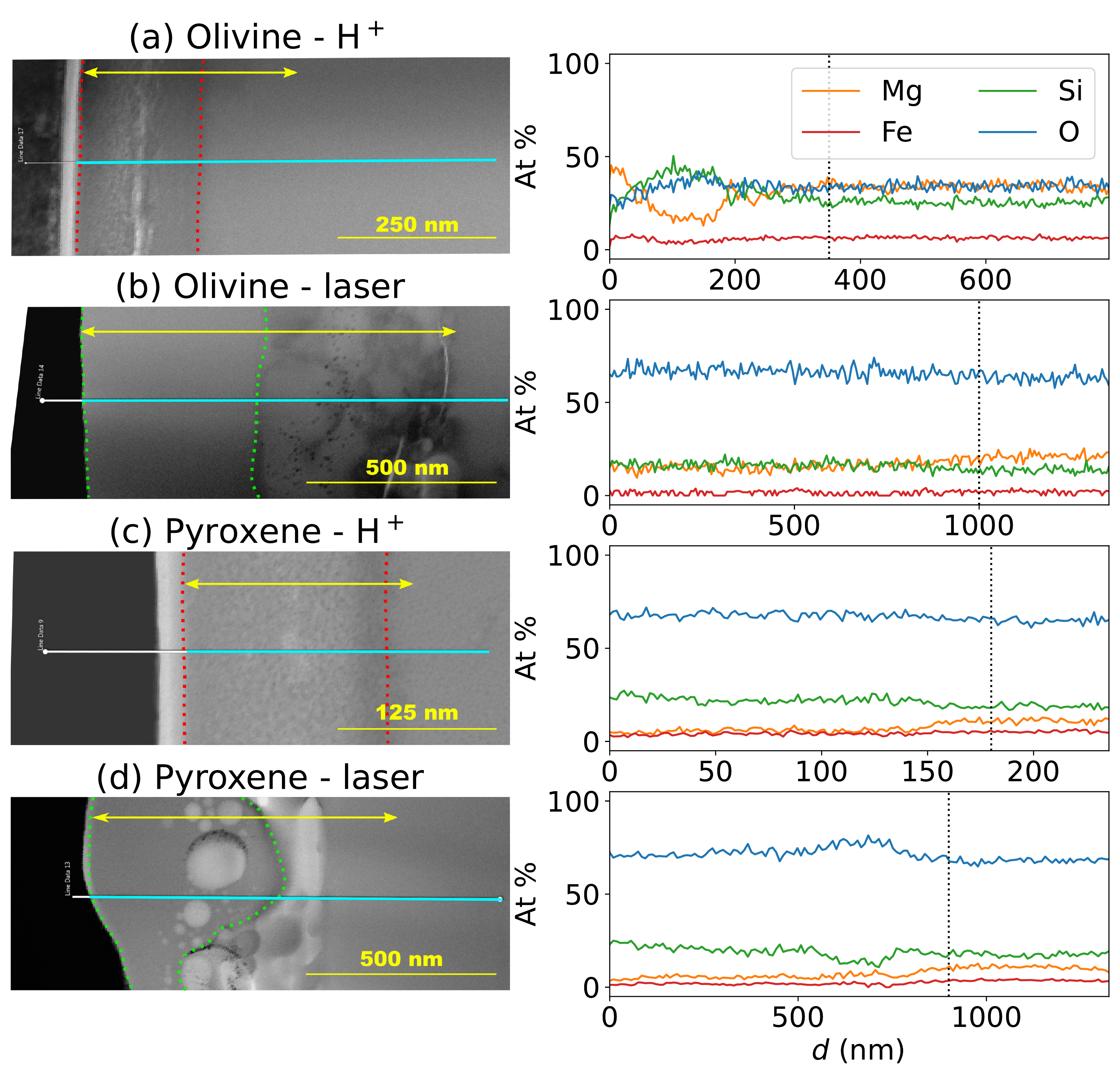}
        \caption{Composition variation in the sub-surface layers. Sample cross-section (left) with the marked (turquoise) position of the composition profile that is shown on the right. The surface of the samples is to the left from the beginning of the turquoise line. The red dashed lines mark the boarders of the regions of partially amorphous layers, and the green dashed lines mark the boarders of the completely amorphous layers. The yellow arrows show an approximate region of chemical changes. (Right) The black dashed lines in the graphs mark an approximate depth, $d$, under which the composition does not change significantly.}
        \label{fig-temchem}
\end{figure}

In Fig.~\ref{fig-temchem}, we show depth profiles of the chemical composition of all four samples. All the samples show a similar change -- depletion of Mg and enrichment of Si within the sub-surface amorphous and partially amorphous zones compared to the un-altered part of the sample. This trend is more pronounced in the OL samples. This happens at $\approx\,250$\,nm for H$^+$ irradiation and at $\approx\,800$\,nm for laser irradiation. The composition is stabilised as we approach underlying unaltered areas. The depths below which the composition does not change significantly are as follows: $\approx\,350$\,nm and 1000\,nm for H$^+$- and laser-irradiated OL, respectively, and $\approx\,180$\,nm and 900\,nm for H$^+$- and laser-irradiated OPX, respectively. In general, the influenced layer in the OL is always slightly thicker than in the OPX irradiated by the same agent. Laser irradiation results in a thicker layer than ion irradiation.

\begin{figure}
        \centering
        \includegraphics[width=\columnwidth]{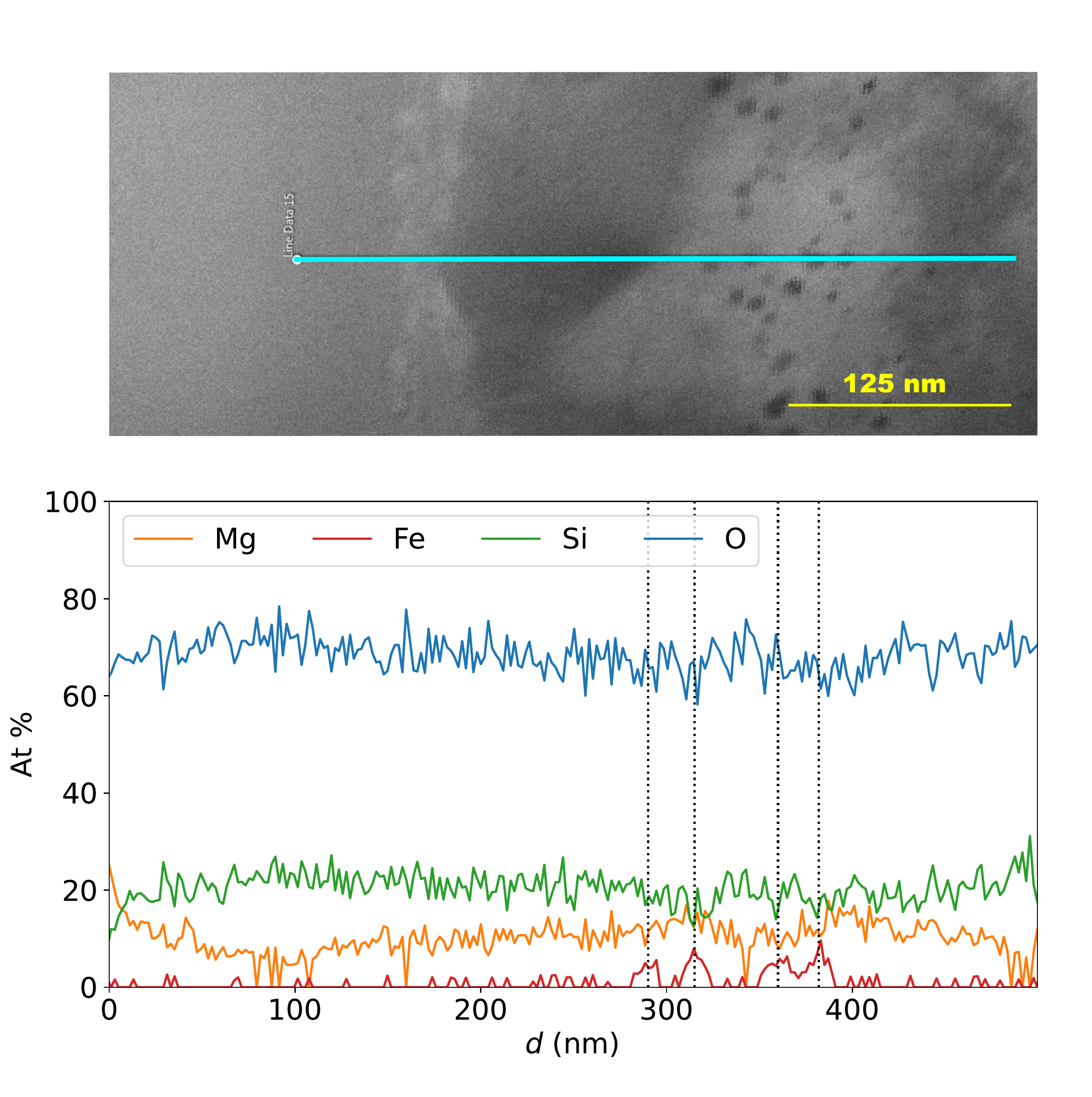}
        \caption{Line scan through the area of nanophase Fe particles. (Top) Bright-field scanning transmission electron microscopy image of an area rich in nanophase Fe particles in the olivine sample irradiated by laser with marked position (turquoise) of the composition profile. (Bottom) Composition profile plot, with the dashed lines indicating approximate locations of nanophase Fe particles.}
        \label{fig-npFe}
\end{figure}

Figure~\ref{fig-npFe} shows a zoomed-in STEM-BF image of the zone containing nanoparticles in the OL irradiated by laser together with a composition profile. The locations of the nanoparticles correlate positively with the abundance of Fe. Also, based on the TEM-EDS point analysis, there is a significant increase in the amount of Fe in the particles, which points to the fact that these are indeed the npFe$^0$ particles.

\section{Discussion}
\subsection{Observed changes}
Our experiments show that the sample surface (viewed from the top) does not exhibit elemental abundance variations. We thus cannot discern between the two weathering agents by observation of changes in elemental abundances on the surface. The immediate sub-surface layers, on the other hand, experience changes in the abundance of some elements. These changes go even deeper into the sample than the amorphous zone. Sudden drops in elemental abundances are indicative of the presence of vesicles (see Fig.~\ref{fig-temchem}).

\begin{figure}
        \centering
        \includegraphics[width=\columnwidth]{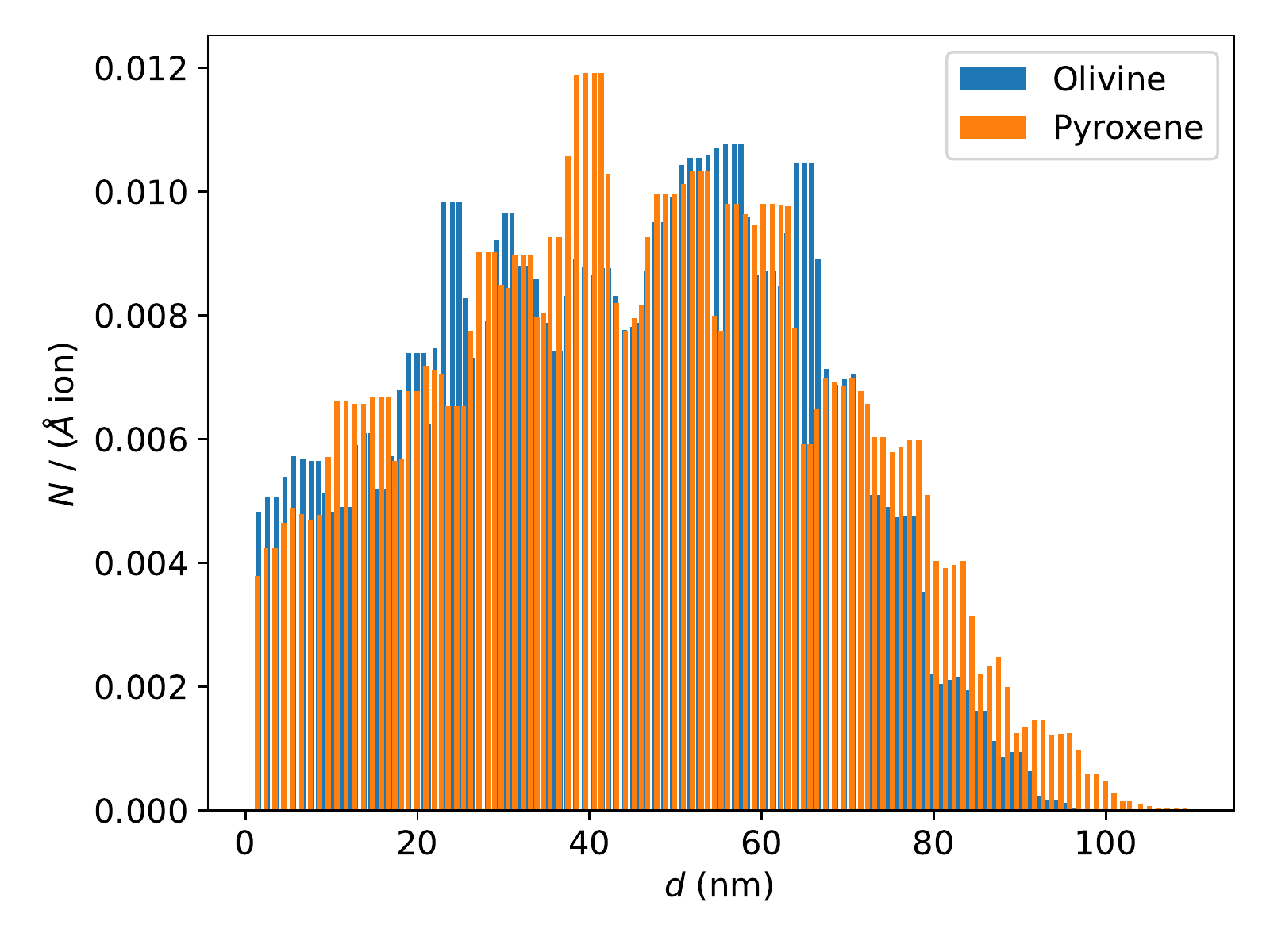}
        \caption{Histogram of collision events in H$^+$-irradiated samples of olivine and pyroxene. $d$ stands for the depth below the sample surface, and $N$ is the number of events.}
        \label{fig-hist}
\end{figure}

We see that there is a difference in the sub-surface structure of H$^+$- and laser-irradiated samples. Those irradiated by H$^+$ mainly show vesicular structures. We notice, as in Fig.~\ref{fig-TEM}~(a, b), that the OL has multiple layers of elongated vesicles, while the OPX shows more spherical and randomly distributed ones. We thus calculated the collisions in the two samples in our experimental conditions (5 keV H$^+$ ions penetrating into a layer of OL or OPX with chemical composition matching our samples) using the SRIM software (see Sect.~\ref{sect-srim}). Based on the SRIM results, we plotted the distributions of the collision events with respect to the depth below the surface for the two samples, as shown in Fig.~\ref{fig-hist}. We can see that both in OL and OPX, the collisions of H$^+$ ions within the sample happen in a very similar range of depths below the surface. These simulations thus do not explain the different distributions of vesicles we observe in our samples. Nevertheless, the SRIM simulations operate in the so-called zero-dose regime. This means that each ion, when impacting the sample in the simulation, behaves as if the sample was un-irradiated. It thus does not take into account the cumulative damage in the sample caused by the previous ions. Olivine is known to be more compliant to maintaining structural changes than pyroxene \citep{quadery_15}. We therefore hypothesise that the collision events in the OL sample were distributed differently to the SRIM simulation as the incoming ions experienced a more altered environment than in the case of OPX, where the vesicles could have distributed more homogeneously.

The penetration depths of H$^+$ ions in OL and OPX, which we obtained from SRIM, agree with the similar thicknesses of the altered layers. In the OL and OPX, we observe that approximately the top 200\,nm and 150\,nm, respectively, are altered. The penetration depths are not significantly different; they are approximately 60\,nm with dispersions of $\approx30$\,nm. 

The changes in the laser-irradiated samples are mainly caused by melting. The OPX sample shows a more global melting pattern spanning across individual laser spots, while in the case of the OL, the melting is rather localised to the area near the laser spot (Fig.~\ref{fig-craters}). This observation agrees with the fact that OPX melts already at lower temperatures than OL. The melting temperature of OL of our composition is approximately 1820$^\circ$C \citep{weizman_97, ctibor_15}, while for OPX it is approximately 1500$^\circ$C \citep{huebner_80}. The OPX thus has a more developed molten surface full of melt drops and melt splashes than the OL sample.

For all the space weathering agents, the altered layers in the OL are thicker than those in the OPX irradiated by the same agent. This is potentially connected to the fact that OPX has a greater ability to heal the defects than OL, as was mentioned by \cite{quadery_15}.

\subsection{Comparison with previous work}
The images of our samples from TEM reveal stratigraphy in the sub-surface layers. Some samples show a purely amorphous layer, such as the laser-irradiated OL; other samples show a partially amorphous layer, such as the H$^+$-irradiated OL. Similar to our findings, \cite{noguchi_14b, noguchi_14a} observed olivine- and pyroxene-rich Itokawa particles and found that their sub-surface structure may be divided into (three) zones. The first zone is composed of sputter and vapour deposits. We observed this only in the laser-irradiated samples. The second zone is composed of partially amorphous material, and the third zone is the underlying unchanged mineral. \cite{weber_20} also observed nano-stratigraphy of the sub-surface layers in their laser-irradiated pressed pellets. In agreement with us, they found that the nano-stratigraphy differs for olivine and pyroxene samples. The thickness of the altered zone observed in our samples is similar to the work of \cite{fazio_18} and \cite{weber_20}.

Our samples contain abundant vesicles, which are small in the H$^+$-irradiated samples and large in the OPX irradiated by laser. \cite{demyk_01} observed vesicles in their samples irradiated by He$^+$ ions as well. The sizes they saw are similar to those in our ion experiment. Even \cite{laczniak_21} observed vesicles in olivine irradiated by H$^+$ and He$^+$ ions. Based on \cite{laczniak_21}, the vesicles are usually smaller than 80\,nm for the Moon, which agrees with our H$^+$ experiments. \cite{kaluna_18} carried out ion irradiations of pristine lunar soils, predominantly made of pyroxene with accessory olivine, and in agreement with us they found vesicles in glassy rims, with sizes similar to our OL case. Olivine- and pyroxene-rich Itokawa particles contain lenticular vesicles that were, based on \cite{matsumoto_15}, created by He$^+$-irradiation. Lenticular vesicles with a longer side parallel to the surface, observed not only by \cite{matsumoto_15} but also \cite{noguchi_14a}, match our H$^+$-irradiated OL observations.

Blisters observed in the H$^+$-irradiated samples are significantly larger than those observed on natural samples. For example, the olivine-rich samples from asteroid Itokawa showed abundant blistering with blister sizes around 30\,nm (varying by a factor of two) \cite[see ][]{dobrica_16}. This may be because Itokawa grains are young and not exposed to the space environment for a long time, as suggested by the presence of completely unblistered regions on the grains (our samples represent approximately 10$^3$\,yr of exposure at 3 au). We only observe blisters in the H$^+$-irradiated samples. In agreement, \cite{matsumoto_18} observed blisters in the surroundings of the micrometeoroid impact craters but never inside them. 

There are several possible reasons for the discrepancy in the blister size between our samples and natural samples. One of them may be the five-times-higher energy of individual H$^+$ ions we used for irradiation compared to the standard solar wind. \citet{demyk_01} used standard solar wind energy of He$^+$ ions and a five-times-higher fluence than in our experiment. As a result, the bubbles they observed did not exceed the size of $\approx400$\,nm. Also, the size of blisters may be influenced by the crystal orientation to the incident angle of the ion beam of low-Ca pyroxene as was pointed out by \citet{matsumoto_15}. Based on \citet{sznajder_18}, the size of blisters is proportional to the irradiation time, which matches the observations of \citet{demyk_01}. We disregard the too high fluence as a reason for the discrepancy as \citet{demyk_01} and \citet{keller_15} used a higher fluence and did not obtain such large blisters. \citet{matsumoto_13}, on the other hand, also used a five-times-higher fluence and obtained blisters as large as 3\,\textmu m (which is larger than blisters in our experiment). However, in their experiments, H$^+_2$ ions were used, not H$^+$, which results in a significant difference.

We observe chemical changes in sub-surface layers in agreement with \cite{laczniak_21}, who observed Mg-depleted, Si-enriched layers in their samples. Their explanation is that this may be due to preferential sputtering of Mg, which leads to non-stochiometric concentrations (see also, for example, \citealt{cassidy_75} and \citealt{lantz_17}). Similarly to us, they found that the chemically distinct layer is thicker than the visually observed one, which may be due to radiation-enhanced diffusion that makes sputtered atoms mobile.

The only sample in which we identify npFe$^0$ particles is the laser-irradiated OL. \cite{weber_20} observed similar behaviour. While the npFe$^0$ particles were observed in olivine, they were sparse and smaller or completely absent from pyroxene in their experiment with nanosecond laser pulses. 

The npFe$^0$ particles we observe are of similar size or smaller than in other experiments or natural samples. Nanoparticles of a similar size have been identified in the laser experiments of \cite{wu_17} or in ion-irradiated pristine lunar soils and mature lunar soils from the Apollo 17 mission \citep{kaluna_18, kling_21}. Based on \cite{noguchi_14b, noguchi_14a}, Itokawa grains contain only small npFe$^0$ particles (5\,nm) induced by solar wind ions, while particles from the Moon have greater nanoparticle size variation \citep{noble_07}. Larger npFe$^0$ particles were observed, for example, by \cite{fazio_18} and \cite{weber_20}. The greater size of nanoparticles may be due to a slightly different set-up of the laser compared with our set-up (e.g. smaller spot size resulting in a higher concentration of energy or greater energy and repeating pulses). As \cite{fulvio_21} noted, the laser spot size significantly influences the resulting craters. Also, the difference in the forsterite content of our samples and samples in \cite{fazio_18} could have played a role. The nature of the sample was different as well; \cite{fazio_18} used a monocrystalline sample, while we used a porous pressed powder surface, which may influence the laser sputtering. \cite{fazio_18} found npFe$^0$ particles both in the glassy layer and in the polycrystalline layer below it. In our case, the majority of the npFe$^0$ particles are situated below the glassy layer. 

\subsubsection{Notes on laser set-up} 
The main difference between (previously widely used) ns laser pulses and fs laser pulses is in the interaction between the laser and the material. Nanosecond pulses mainly cause heating and melting of a target. In this regime, a thermal wave propagates through the material, which results in a significant and widespread layer of melt. In the fs regime, on the other hand, the thermal conduction to the target may be, in the first approximation, neglected \citep{chichkov_96}. Because of the short timescale of the pulse, the interaction proceeds in a different way. The material may undergo fragmentation and immediate transformation from solid to vapour. The brevity of the pulse also disables the interaction of the laser with the vapour plume. At high peak irradiance, the material undergoes melting, which is only confined to the area of the laser spot, which contrasts with the ns regime \citep{fazio_18, perez_03}.

The wavelength of the laser pulse we used (800\,nm) is different to the usually applied Nd-YAG or ultraviolet (UV) lasers (1064\,nm or 248\,nm see, for example, \citet{sasaki_01a, brunetto_06, matsuoka_15}. As \citet{brunetto_06} noted, the 1064-nm pulses cause the laser to behave as a heat source, resulting in uncontrolled melting. The UV pulses, on the other hand, cause direct breaking of the bonds in the target with little damage to the surrounding material. The wavelength of our laser experiment is closer to the 1064-nm than to the UV experimental set-up. Also, both, the 800 and 1064\,nm lasers operate at the wavelengths matching the main absorptions in the olivine (and pyroxene) materials. Nevertheless, the evaluation of the differences in the influence of the two set-ups on the final state of the material is beyond the scope of this article, and we refer the reader to the original publication \citep{fazio_18} for a deeper discussion of the experimental set-up.
 
\subsection{Sub-surface structure and reflectance spectra}
While the OPX irradiated by laser shows a major layer of amorphous material, for all the other types of irradiation, the sub-surface alteration does not exceed 300\,nm. If we compare this information with the spectral evolution (Fig.~\ref{fig-spectra}), we see that at shorter wavelengths the thickness of the altered layer is similar to the wavelength of light. With longer wavelengths, this does not hold. The incident light at longer wavelengths probes more deeply into unaltered layers, and the spectral change is thus smaller as the spectrum represents a mixture of the fresh and mature material.

The H$^+$ irradiation causes a similar sub-surface change in the OL and OPX sample (Fig.~\ref{fig-TEM}), that is partial amorphisation of topmost 200\,nm and 150\,nm, respectively. The subsequent spectral change is characterised by a strong decrease in reflectance in the VIS and nearly no change in the NIR region. This darkening may be explained by a change in the imaginary part of the refractive index after amorphisation that can be of up to several orders of magnitude (see, for example, olivine refractive indices in \cite{dorschner_95} and \cite{zeidler_11}). The absorption in the volume is a function of the thickness of the volume compared to the wavelength. As the thickness of the amorphous layer is small (a few hundred nanometres), the VIS wavelengths are absorbed much more than the longer wavelengths. The partially amorphous layer is thicker in the OL sample, which causes the greater change in the spectrum (see Fig.~\ref{fig-spectra}). The small vesicles in the samples probably cause scattering, which may mildly increase the brightness. Nevertheless, this effect seems to be minor compared to the darkening. 

The OL irradiated by laser shows a similar spectral change to the one irradiated by H$^+$, but the absolute value of the change differs. The reason for this may be that the amorphous layer is thicker and the amorphisation in the laser-irradiated sample is complete compared to the partial amorphisation in the ion irradiation case. Also, the laser-irradiated samples show a molten surface, which is smoother than the original surface. As a result, the spectra darken because of the smaller contribution of the reflection from the rough surface. Laser irradiation of OL causes the additional creation of npFe$^0$ particles. Their presence also causes a slight change in the NIR spectral slope, as observed in previous work \citep[see, for example, ][]{pieters_16}. 

The OPX irradiated by laser shows a different sub-surface structure to the OL, the main features being a thick glassy layer and plenty of large (tens to hundreds of nm) vesicles. The light scatters on the vesicles, which decreases the effect of darkening compared to the other cases. On the other hand, we hypothesise that the significant amorphous layer makes the material in the top layers compact, causing that the scattering on the micro-roughness is disabled. Due to the presence of large vesicles, the light penetrates deep into the material, where the volume scattering and absorption are enhanced, which may hypothetically increase the absorption bands' depths compared to the pristine sample, where surface scattering on the individual grains diminishes the contribution of the volume absorption. All these observations are summarised in Table~\ref{tab-explain2}.
\begin{table*}
        \caption{Sub-surface structure for different materials and irradiation sources.}
        \label{tab-explain2}
        \centering
        \begin{tabular}{c c c c c}
                \hline \hline
                & npFe$^0$ & Vesicles & Amorphization & $t$ (nm)\\
                \hline
                OL -- H$^+$ & no & yes  &  partial & 200 \\
                OL -- laser & yes & no  &  full & 300 \\
                OPX -- H$^+$ & no & yes &  partial & 150 \\
                OPX -- laser & no & yes &  full & 400 -- 2500 \\
                \hline
        \end{tabular}
        \tablefoot{OL stands for olivine, OPX for pyroxene, npFe$^0$ for nanophase Fe particles, and $t$ for thickness of the altered layer. The individual columns state whether we identified the object in our samples or not.}
\end{table*}

\subsection{Application to Solar System research}
Based on our findings we can make the following prediction of the sub-surface and spectral evolution due to space weathering. In the first $10^3$--$10^4$\,yr after exposure of the material to the space environment, the solar wind dominates \citep[e.g. ][]{blewett_11} and the airless planetary surfaces will mainly suffer from partial amorphisation, which will induce rapid changes in the VIS spectral slope. Only later will small ($<$ 5\,nm) npFe$^0$ particles appear, as we now see in Itokawa particles brought by the Hayabusa mission, which will also mildly influence the NIR slope. On timescales longer than $10^8$\,yr, we expect micrometeoroid bombardment to become significant \citep{blewett_11, noguchi_11}, which may create larger npFe$^0$ particles (as seen in the OL sample) and/or large vesicles in thick amorphous layers (as seen in the OPX sample). The OPX changes will not influence spectral slope any more but will have an effect on absorption bands, as we observe in asteroids (433)~Eros and (4)~Vesta.

\section{Conclusions}
Our analyses revealed a significant difference in the morphology of laser- and ion-irradiated surfaces. While ions (simulating the influence of solar wind) induced the creation of blisters, laser shots (representing micrometeoroid impacts) caused melting with the associated creation of bubbles and melt splashes. We did not identify any spatial variation in chemical abundances of the surfaces irradiated by laser and by H$^+$, which means that the two space weathering agents are not distinguishable by this analysis. 

Samples irradiated by H$^+$ showed a very similar sub-surface structure rich in small vesicles in thin (up to 200\,nm), partially amorphous layers. The laser-irradiated OL showed a fully amorphous layer below which we identified npFe$^0$ particles. The laser-irradiated OPX contained, unlike the previous cases, a thick amorphous layer full of large vesicles. 

Based on a comparison of the irradiation-induced structures and the changes in reflectance spectra, we found that the thickness and the structure of the amorphous layer have a major influence on the spectra. In both materials, H$^+$ irradiation resulted mainly in VIS spectral slope changes due to the thickness of partially amorphous layers being comparable to VIS wavelengths. Spectral slope changes throughout the VIS-NIR wavelengths observed in the olivine sample irradiated by laser resulted from the presence of thicker, completely amorphous layers together with the presence of the npFe$^0$ particles. The pyroxene sample irradiated by laser showed the thickest, entirely amorphous layer that, in combination with large vesicles, probably caused changes of the absorption bands' depths and no continuum changes. 

We may thus predict the sub-surface and spectral evolution of material exposed to the space environment with time: initially thin, partially amorphous layers with a thickness comparable to the wavelength of observation will cause significant changes to the VIS spectral slope. The later appearance of npFe$^0$ particles will also influence the NIR slope, and a subsequent increase in the thickness of the amorphous layer and vesicle size will induce only mild darkening and modification of depths of the absorption bands.

\begin{acknowledgements}
      The authors would like to express many thanks to the anonymous referee and to the editor for thorough corrections and comments on this manuscript.
      
      This work was supported by the University of Helsinki Foundation and the Academy of Finland project nos 325805, 1335595 and 293975, and it was conducted with institutional support RVO 67985831 from the Institute of Geology of the Czech Academy of Sciences. The authors acknowledge funding from Charles University (Project Progres Q47). This work has been realized within the research infrastructure SUSEN established in the framework of the European Regional Development Fund in project CZ.1.05/2.1.00/03.0108 and of the European Structural and Investment Funds in project CZ.02.1.01/0.0/0.0/15\_008/0000293.
\end{acknowledgements}

\bibliographystyle{aa} 
\bibliography{literature} 


\end{document}